\documentclass[preprint,showpacs,nofootinbib,preprintnumbers,amsmath,amssymb]{revtex4}

\usepackage{graphicx}
\usepackage{dcolumn}
\usepackage{bm}

\newcommand{\be}{\begin{equation}}
\newcommand{\ee}{\end{equation}}
\newcommand{\ZZ}{{\sf Z}}
\newcommand{\comment}[1]{}

\begin{document}


\title{
Non-commutative Field Theory with Twistor-like Coordinates}

\author{Tomasz R. Taylor}
\affiliation{
Department of Physics, Northeastern University\\
Boston, MA 02115, United States of America
}%

\begin{abstract}We consider quantum field theory in four-dimensional Minkowski spacetime, with the position coordinates represented by twistors instead of the usual world-vectors.
Upon imposing canonical commutation relations between twistors and dual twistors, quantum theory of fields described by non-holomorphic functions of twistor variables becomes manifestly non-commutative, with Lorentz symmetry broken by a time-like vector.
We discuss the free field propagation and its impact on the short- and long-distance behavior of physical amplitudes in perturbation theory. In the ultraviolet limit, quantum field theories in twistor space are generically less divergent than their commutative counterparts. Furthermore, there is no  infrared--ultraviolet mixing problem.
\end{abstract}

\maketitle
Twistor theory \cite{penrose} offers an alternative description of the four-dimensional spacetime in which spinors are regarded more fundamental than world-vectors. At the classical level, twistors provide very convenient tools for analyzing certain aspects of four-dimensional ge{\nolinebreak}ometry \cite{penrin}. Twistor quantization \cite{Penrose:1968me,Penrose:1972ia,penrose} leads, however, to a radical departure from the concept of spacetime continuum. It makes points fuzzy while maintaining a well-defined concept of null direction, as opposed to
the conventional spacetime quantization that keeps  points well-defined, but with {\it light cones\/} fuzzy due to the fluctuating metric. Thus the effects of twistor quantization are {\it not\/} limited to quantum gravity: they affect quantization of all fields. Most of previous studies of quantum fields in twistor space \cite{Penrose:1972ia,penrose} have been limited to holomorphic fields that are in some way insensitive to twistor quantization. Here, we consider the more general case of non-holomorphic fields. As explained below, their dynamics
are indeed affected by the non-commutative structure of  quantum twistor spacetime.

The twistor coordinates in Minkowski space are pairs of two-component spinor fields
\be\ZZ^{\alpha}=(\omega^{A}\, ,\,{\pi}_{A'}) ~;~ \alpha=1,2,3,4,~~ A,A'=1,2,\label{twist}\ee with
\be \omega^{A}=\omega^{\hskip -1.8mm ^{\circ}\;A }-ix^{AA'}\!
\pi^{{\hskip -1.7mm} ^{\circ}}_{A'}
\quad ,\quad {\pi}_{A'}=\pi^{{\hskip -1.7mm} ^{\circ}}_{A'}~,\label{om}\ee
where the hermitean matrix $x^{AA'}$ represents the position vector and $(\omega^{\hskip -1.8mm ^{\circ}\;A }, \pi^{{\hskip -1.7mm} ^{\circ}}_{A'} )$ are constant spinors whose values coincide with those of $(\omega^{A},{\pi}_{A'})$ at the origin.\footnote{We use the notation and conventions of Penrose and
Rindler, Ref.[2]. In particular, $x^{00'}=(t+z)/\sqrt{2}$ and the
speed of light $c=1$. The spinor indices are raised and lowered as
usual, by the two-index antisymmetric tensor $\epsilon$: for
instance, $ \psi_A\to\psi^A=\epsilon^{AB}\psi_B$.}
\comment{We define
\be \ZZ^{\hskip -2.1mm ^{{^\circ}}\,\alpha}=(\omega^{\hskip -2.1mm ^{\circ}\;A }, \pi^{{\hskip -2.0mm} ^{\circ}}_{A'} )\quad,\quad  z^{\alpha }=(-ix^{AA'}\!
\pi^{{\hskip -2.0mm} ^{\circ}}_{A'} ,0 )\quad\Rightarrow\quad \ZZ^{\alpha}= \ZZ^{\hskip -2.1mm ^{{^\circ}}\,\alpha}+z^{\alpha}  \ee}
By complex conjugation, one obtains the dual twistor
\be\bar\ZZ_{\alpha}=({\bar\pi}_{A}\, ,\,{\bar\omega}^{A'}),\ee
where ${\bar\omega}^{A'}=\omega^{A\, *}$ and ${\bar\pi}_{A}={\pi}_{A'}^{\, *}$. In quantum theory, the twistors $\ZZ^{\alpha}$ and $\bar\ZZ_{\alpha}$ become operators satisfying the canonical commutation relation \cite{Penrose:1968me,Penrose:1972ia}
\be [\,\ZZ^{\alpha},\bar\ZZ_{\beta}\,]=\hbar\delta^{\alpha}_{\beta}\, .\label{commutator}\ee

The fields on twistor space are usually assumed to be holomorphic functions of twistor variables \cite{Penrose:1972ia,penrose} which, according to Eq.(\ref{commutator}), remain commuting upon quantization. They can be mapped into the functions satisfying the conventional field equation in Minkowski space (like $\Box\phi=0$) by using the celebrated Penrose transform \cite{penrin,penrose,ptran}. With such a restriction, however, it is difficult to construct a quantum field theory in twistor space that would be related in a more or less straightforward way to the second quantization in Minkowski space.\footnote{A particularly interesting idea in this
direction is to study the so-called twistor diagrams, see {\it
e.g}.\ Refs. \cite{Penrose:1972ia,penrose,diagrams}. It seems very difficult
though, to make a connection between twistor diagrams and Lagrangian
formalism. One of the problems is that even the standard plane wave
functions have no simple representation in terms of holomorphic
twistor integrals.}

The null twistors, satisfying $\ZZ^{\alpha}\bar\ZZ_{\alpha}=0$, represent null geodesics in Minkowski space, generated by $x^{AA'}\to x^{AA'}+k\,\bar\pi^{{\hskip -1.7mm} ^{\circ}\,A}\pi^{{\hskip -1.7mm} ^{\circ}\,A'}$ with real $k$ \cite{penrin}. In  order to represent points as intersections of such geodesics, one has to introduce two orthogonal, null twistors $\ZZ^{\alpha}_a$, $a=1,2,$ \cite{penrose,penrin} such that
\be  \ZZ^{\alpha}_a\bar\ZZ_{\alpha\, b}=\omega^A_a \bar\pi_{bA}+
 \pi_{aA'}\bar\omega^{A' }_b=0\quad ;\quad a,b=1,2.\label{null}\ee
\comment{{}For any two twistors associated to the same point $x$, of the form given in Eq.(\ref{twist}), the above condition is equivalent to
\be \ZZ^{\hskip -2.1mm ^{{^\circ}}\,\alpha}_a\bar\ZZ^{\hskip -2.1mm ^{{^\circ}}}_{\alpha\, b}=\omega^{\hskip -2.1mm ^{\circ}\;A }_a \bar\pi^{{\hskip -2.0mm} ^{\circ}}_{bA}-
\bar\omega^{\hskip -2.1mm ^{\circ}\;A' }_b \pi^{{\hskip -2.0mm} ^{\circ}}_{aA'}=0.\ee}
Then the position of the point is
\be x^{AA'}=\frac{i}{\pi_{1B'}\pi_2^{B'}}[(\omega^{A}_1-\omega^{\hskip -1.7mm ^{\circ}\;A }_1)\pi^{A'}_2-(\omega^{A}_2-\omega^{\hskip -1.7mm ^{\circ}\;A }_2)\pi^{A'}_1].\label{xterm}\ee
Here, we expressed $x$ in terms of  the holomorphic twistor variables. Of course a similar formula holds for $x$ expressed in terms of the dual (anti-holomorphic) variables. Both descriptions are equivalent (and consistent with each other) due to Eq.(\ref{null}). We adopt the point of view that twistor coordinates are fundamental, thus vectors should merely appear as auxiliary coordinates.
\comment{However, even after imposing Eq.(\ref{null}), the space of ``bitwistors'' parameterized by $(\ZZ_1, \ZZ_2)$ has six complex dimensions, therefore we have to decide first how to embed the {\it real\/} Minkowski spacetime in such a higher-dimensional space.}
In order to parameterize Minkowski space, we freeze the $\pi$-components and use the $\omega$-components as the primary coordinate variables. Note that even with fixed $\pi_{a A'}=\pi^{{\hskip -1.7mm} ^{\circ}}_{aA'}$, Lorentz symmetry can be preserved by assuming that the internal indices labelling the two twistors transform in a way that compensates the $SL(2,\mathbb{C})$ transformations acting on the spinor indices.

We quantize twistors by generalizing the commutator (\ref{commutator}) to
\be [\,\ZZ_a^{\alpha},\bar\ZZ_{b\beta}\,]=\hbar\delta^{\alpha}_{\beta}\delta_{ab}\, .\label{newcom}\ee
Now the symmetry group acting on the internal indices is limited to $U(2)$, hence freezing the $\pi$-components breaks Lorentz symmetry down to the $SO(3)$ subgroup of spatial rotations.
The order parameter is the time-like vector
\be\label{order}l_{AA'}=\bar\pi_{1A}\pi_{1A'}+\bar\pi_{2A}\pi_{2A'} ~,~ l^2=2|\pi_{1A'}\pi_2^{A'}|^2\equiv\mu^2 ~.\ee

Translations of the holomorphic functions of $\ZZ_1, \ZZ_2$ are generated by
\be \frac{\partial}{\partial x^{AA'}}=-i\,\pi_{aA'}\frac{\partial}{\partial\omega_a^A}\ee
Thus the momentum operator, restricted to the subspace of holomorphic functions is
\be p_{AA'}=-\hbar\,\pi_{aA'}\frac{\partial}{\partial\omega_{a}^A} ~.\ee
Its eigenfunctions are
\be f_p(\omega)=e^{-\omega_a^A\tilde\omega_A^{\,a}/\hbar}\quad,\quad p_{AA'}=\tilde\omega_A^{\, a}\pi_{aA'} ~.\ee
Of course, the spinors $\tilde\omega$ are must satisfy the reality condition
\be\label{real} \tilde\omega_A^{\, a}\pi_{aA'} =\bar\pi_{aA}\bar{\tilde\omega}_{A'}^{\, a} ~.\ee

We are interested in non-holomorphic fields depending on the coordinates $\omega$ and ${\bar\omega}$. The commutation relations (\ref{newcom}) ensure that these are commuting variables as long as they are associated to the same point $x$. In particular,
\be [\omega^{{\hskip -1.7mm}^{\circ}\,A}_a,\bar\omega^{{\hskip -1.7mm} ^{\circ}\,A'}_b]=
[\pi^{{\hskip -1.7mm}^{\circ}}_{aA'},\bar\pi^{{\hskip -1.7mm} ^{\circ}}_{bA}]
=0\, .
\ee
However, the non-vanishing commutators
\be [\omega^{{\hskip -1.7mm}^{\circ}\,A}_a,\bar\pi^{{\hskip -1.7mm} ^{\circ}}_{bB}  ]=\hbar\delta^A_B\delta_{ab}\quad,\quad    [\pi^{{\hskip -1.7mm}^{\circ}}_{aA'},\bar\omega^{{\hskip -1.7mm} ^{\circ}\,B'}_b]=\hbar\delta^{B'}_{A'}\delta_{ab}  ~,\ee
combined with Eq.(\ref{om}), imply that for the coordinates
$\omega^{A}_{a}(x_1)$ and ${\bar\omega}^{A'}_{b}(x_2)$ associated to two arbitrary points $x_1$ and $x_2$, respectively,
\be \label{omcom}[\omega^{A}_{a}(x_1),{\bar\omega}^{A'}_{b}(x_2)]=i\hbar (x^{AA'}_2-\, x^{AA'}_1)\delta_{ab}\, .\ee
Thus while the coordinates $\omega$ and ${\bar\omega}$ commute {\it locally}, they are {\it non-locally non-commutative}. It is a ``foggy'' type of commutator, with the uncertainty growing with separation. Other commutation relations of the  full quantum twistor algebra are:
\begin{eqnarray} [\omega^{A}_a(x),\bar\pi_{bB}  ] = \hbar\delta^A_B\delta_{ab}\, ,& [\pi_{aA'},\bar\omega^{B'}_b(x)]=\hbar\delta^{B'}_{A'}\delta_{ab}\, ,\hskip 2mm \nonumber \\[-1mm]\label{othcom}\\[-1mm]
[\omega^{A}_a(x_1),\omega^{B}_b(x_2)]=[\pi_{aA'} ,\hskip -4mm&\pi_{bB'}]=
[\pi_{aA'},\bar\pi_{bA}]
=0\, .\nonumber
\end{eqnarray}
Note that the positions $x$ defined in Eq.(\ref{xterm}) are c-number quantities commuting with all twistor variables. It is easy to show that Eq.(\ref{xterm}) is consistent with Eqs.(\ref{omcom},\ref{othcom}) and to check that the algebra is closed.

As already announced before, we will be considering fields with the locally commuting
$\omega$-components as the position coordinates. The step of choosing a fixed $\pi$ should be then regarded as the selection of one specific field mode. Due to the reality condition (\ref{null}), a given $\pi$ defines Minkowski spacetime as a (pseudo-real) hypersurface of the complex $\omega$-space. The associated field mode propagates on the corresponding hypersurface. In that respect, our use of coordinates is different from traditional twistor theory because it is based on a direct map from twistors to Minkowski spacetime instead of the Penrose transform \cite{ptran}. Since Lorentz invariance is broken by the choice of the hypersurface, {\it c.f}.\ Eq.(\ref{order}), it is appropriate to refer to the spacetime parameterized by $\omega$ as to {\it foggy {\ae}ther}.
In the rest of this article, we study some properties of quantum fields propagating in foggy {\ae}ther.

Let us consider a free, real massless scalar field
\be \label{scalar} \phi(\omega,\bar\omega)=\int\!\!\frac{d^3\vec{p}}{\sqrt{(2\pi)^32|\vec{p}\,|}}
\left(a_pe^{\omega_a^A\tilde\omega_A^{\,a}/\hbar}+a^{\dagger}_pe^{\bar{\tilde
\omega}_{A'}^{\,a}\bar\omega_a^{A'}\!\!/\hbar}\right).\ee
Here, $\tilde\omega$ represent momenta satisfying the on-shell condition $p^2=0$, hence they can be parameterized as $(\tilde\omega_A^{1},\tilde\omega_A^{2})=(\tilde\rho_A, \lambda\tilde\rho_A)$, with $\lambda$ and $\tilde\rho_A$ subject to the reality condition following from Eq.(\ref{real}).
The momentum integration measure can be written in terms of such twistor variables,  however, for practical purposes, it is more convenient to use the standard ${d^3\vec{p}}$ measure. The creation and annihilation operators satisfy the canonical commutation relation
\be\label{acom}[a_p,a_{p'}^{\dagger}]=\delta^3(\vec{p}-\vec{p}\,') ~.\ee

We are interested in the free Feynman propagator
\be iD(x'-x)=\langle 0|\phi(\omega',\bar\omega')\phi(\omega,\bar\omega)|0\rangle\theta(t'-t) +\langle 0|\phi(\omega,\bar\omega)\phi(\omega',\bar\omega')|0\rangle\theta(t-t'),\label{fey}\ee
where $\omega\equiv\omega(x)$ and $\omega'\equiv\omega(x')$. The Heaviside (step) $\theta$-functions  enforce the time-ordering of field operators.
We take the product of free fields of Eq.(\ref{scalar})
and apply the Baker-Hausdorff formula together with the commutator given in Eq.(\ref{omcom}):
\be\label{baker}   e^{\omega_a^A(x')\tilde\omega_A^{\,a}(p')/\hbar}\,e^{\bar{\tilde
\omega}_{A'}^{\,b}(p)\bar\omega_b^{A'}(x)/\hbar} =e^{[\omega_a^A(x')\tilde\omega_A^{\,a}(p')+\bar{\tilde
\omega}_{A'}^{\,a}(p)\bar\omega_a^{A'}(x)+\frac{i}{2}(x-x')^{AA'}\tilde\omega_A^{\,a}(p')\bar{\tilde
\omega}_{A'}^{\,a}(p)]/\hbar}\ee
After using Eq.(\ref{acom}), we obtain
\be\label{feyres} iD(x'-x)=\int\!\!\frac{d^3\vec{p}}{{(2\pi)^32|\vec{p}\,|}}\left[e^{-ip\cdot (x'-x)(p+2l)^2/(4\mu^2\hbar)}\theta(t'-t)+e^{ip\cdot (x'-x)(p+2l)^2/(4\mu^2\hbar)}\theta(t-t')\right],\ee
where $l$ is the time-like vector introduced in Eq.(\ref{order}). We integrate over three-momenta in the isotropic {\ae}ther frame with
\be \label{frame} l=(\mu, \vec{0}\, )\quad ,\quad x'-x=(t,\vec{r}\, )~.\ee
The result is
\be\label{result} D(t,\vec{r}\,)=\theta(t)[{\cal E}(t+r)-{\cal E}(t-r)]-\theta(-t)[{\cal E}(-t-r)-{\cal E}(-t+r)]~,\ee
where
\be\label{efun}{\cal E}(\tau)=\frac{\hbar}{4\pi^2r}\int_0^{\infty}\frac{e^{-iE\tau/\hbar}\, dE}{(\sqrt{1+8E/\mu}+1)\sqrt{1+8E/\mu}}~.\ee

In the formal limit of infinite mass scale $\mu$,
\be\label{liml}{\cal E}(\tau)\xrightarrow[\mu\to\infty]{}\frac{\hbar^2}{8\pi r}[\delta (\tau)-P\frac{i}{\pi\tau}]~,\ee
where $P$ denotes the principal value, thus the result (\ref{result}) yields the standard
Lorentz-invariant propagator of a massless scalar field in Minkowski spacetime. As expected, non-commutative effects disappear in the infrared limit, for a field propagating over large distances. On the other hand, short-distance propagation in foggy {\ae}ther
is very different from the propagation in the usual relativistic vacuum. If the zero-distance limit is taken with $r\equiv \epsilon_r\to 0 $ first and then $t\equiv\epsilon_t\to 0$, the standard Feynman propagator diverges quadratically as $(\epsilon_r\epsilon_t)^{-1}$, while the propagator of Eqs.(\ref{result}) and (\ref{efun}) exhibits only a milder, $\epsilon_t^{-1}$ linear singularity. With the limit taken in the opposite order, one finds $\epsilon_r^{-2}$ {\it vs.\/} $\epsilon_r^{-1}$. {}Furthermore, the light-cone singularity at $t=\pm r+\epsilon$, $\epsilon\to 0$, which yields a linear divergence $\epsilon^{-1}$ in the standard case, is now replaced by $\ln\epsilon$.
To summarize, the non-commuting property (\ref{omcom}) of twistor coordinates leads to a softer behavior of the Feynman propagator in the ultraviolet limit, without affecting its infrared behavior.

The propagator (\ref{result}) can be rewritten in terms of its Fourier transform as:\be
D(x'-x)=\int\!\!\frac{d^4k}{(2\pi)^4}\, e^{-ik(x'-x)/\hbar}\widetilde{D}(k),\ee
where \be\label{dofk}
\widetilde{D}(k)=\frac{1}{k^2+i\varepsilon}\times 2 \left[\sqrt{8|\vec{k}|/\mu+1}(\sqrt{8|\vec{k}|/\mu+1}+1)\right]^{-1},\ee
with Feynman's $i\varepsilon$ prescription. The milder character of short-distance singularities manifests now in a stronger suppression of high-momentum modes:  $\widetilde{D}(k)\sim |\vec{k}|^{-3}$ for $|\vec{k}|\gg\mu$. As mentioned before, non-commutative effects disappear in the infrared limit: $\widetilde{D}(k)\sim k^{-2}$ at $|\vec{k}|\ll\mu$. Note that the
momentum propagator (\ref{dofk}) contains only one singularity, a pole at $k^2=0$, which reflects the fact that non-commutativity does not affect the standard relativistic dispersion relation.

In non-commutative field theories, the interactions are usually modified compared to the commuting case. For example, on non-commutative Groenewold-Moyal plane, local field products are replaced by the so-called star products \cite{Douglas:2001ba}. Foggy {\ae}ther is different in that respect because it is {\it locally} commutative, therefore local interactions need no modifications.
Thus at least in perturbation theory, the modifications of tree-level amplitudes and of their loop corrections (Feynman diagrams) are completely determined by the prop{\nolinebreak}erties of Feynman propagators described in the previous paragraph. It is clear from our discussion that non-commutative field theories in twistor space do {\it not\/} suffer from the notorious infrared--ultraviolet mixing problem
\cite{Minwalla:1999px,Grosse:2004yu}. Furthermore, the ultraviolet divergences of loop diagrams are milder than those of
their commutative counterparts. {}For instance, na\"ive power counting, based on Eq.(\ref{dofk}), indicates that foggy gauge theories are perturbatively finite. More thorough analysis is necessary, however, to confirm that this is indeed the case.

One of the most striking features of non-commutative field theory discussed in this paper is the violation of Lorentz invariance. It is natural to identify the isotropic {\ae}ther frame (\ref{frame}) with the frame of the cosmic microwave background. Then in Earth-based experiments not only boost-symmetry would be  vi{\nolinebreak}olated, but also spatial anisotropy could be observable.\footnote{I am grateful to Alan Kosteleck\'y for very useful
correspondence.} The non-commutativity mass scale $\mu$ must be sufficiently high, certainly above 10 TeV \cite{Carroll:2001ws}, in order to comply with all experimental constraints. The existing data, see {\it e.g}.\ Ref.\cite{Cane:2003wp}, provide very stringent bounds on various Lorentz symmetry vi{\nolinebreak}olating parameters \cite{cck}.  Without addressing the question how these parameters are related to $\mu$, it is premature, however, to
quote here some stronger bounds. We leave detailed analysis of Lorentz symmetry violation to a future study.

This material is based upon work supported by the National Science Foundation Grant No.\ 0600304. Any opinions, findings, and conclusions or
recommendations expressed in this material are those of the author and do not necessarily
reflect the views of the National Science Foundation.

\bibliography{twist1}

\end{document}